\documentclass[conference]{IEEEtran}
\IEEEoverridecommandlockouts
\usepackage{cite}
\usepackage{amsmath,amssymb,amsfonts}
\usepackage{algorithmic}
\usepackage{graphicx}
\usepackage{graphicx,epstopdf}
\usepackage[table]{xcolor}
\usepackage{textcomp}
\usepackage{xcolor, soul}
\usepackage{multicol}
\usepackage{orcidlink}
\hypersetup{hidelinks}
\usepackage{comment}
\usepackage{booktabs}
\usepackage{balance}
\def\BibTeX{{\rm B\kern-.05em{\sc i\kern-.025em b}\kern-.08em
    T\kern-.1667em\lower.7ex\hbox{E}\kern-.125emX}}

\begin{document}
\title{%An Ensemble-based Kolmogorov Arnold Network for Intrusion Detection Systems in IoT Environment.
Enhancing Intrusion Detection in IoT Environments: An Advanced Ensemble Approach Using Kolmogorov-Arnold Networks
%\thanks{This work is done in the Tecore Networks Lab at Florida Atlantic University and is funded by the Office of the Secretary of Defense (OSD), Grant Number W911NF2010300.}
}

\author{
  Amar Amouri\\
  Abu Dhabi Polytechnic,  Abu Dhabi, UAE \\
%  \textit{amar.amouri@actvetgov.ae} \\
  \thanks{Corresponding author: Amar Amouri\{\textit{amar.amouri@actvetgov.ae}\} } \\
  \and
  Mohamad Mahmoud Al Rahhal, and Yakoub Bazi \\
  College of Computer \& Information Science \\
  King Saud University,   Riyadh, Saudi Arabia \\
  %\textit{mmalrahhal@ksu.edu.sa, ybazi@ksu.edu.sa} \\
  \and
  % Yakoub Bazi \\
  % College of Computer \& Information Science \\
  % King Saud University \\
  % Riyadh, Saudi Arabia \\
  % \texttt{ybazi@ksu.edu.sa}

% \IEEEauthorblockN{Ismail Butun \orcidauthorA{}, and Imad Mahgoub \orcidauthorB \thanks{Corresponding author: Imad Mahgoub \textit{mahgoubi@fau.edu}} }
Ismail Butun, %\orcidauthorA{}
and Imad Mahgoub %\orcidauthorB 
\\
% \thanks{Corresponding author: Imad Mahgoub \textit{mahgoubi@fau.edu}} 
%}
% \IEEEauthorblockA{\textit{Department of Electrical Engineering and Computer Science (EECS),} \\
% \textit{Florida Atlantic University,} Boca Raton, FL, USA; emails:\textit{ibutun@fau.edu}, \textit{mahgoubi@fau.edu}
Department of Electrical Engineering and Computer Science,
Florida Atlantic University, Boca Raton, FL, USA \\ 
%\textit{ibutun@fau.edu}, \textit{mahgoubi@fau.edu}
}

%\and

% \and

% % for copyright statement
% \IEEEoverridecommandlockouts

% % For IEEE CopyRight Statement!
% \IEEEpubid{\makebox[\columnwidth]%{U.S. Government work not protected by U.S. copyright \hfill}
% {979-8-3503-6491-0/24/\$31.00 ©2024 IEEE \hfill}
% %{978-1-5386-5541-2/18/\$31.00~\copyright2018 IEEE \hfill} 
% \hspace{\columnsep}\makebox[\columnwidth]{ }}

\maketitle

%% for copyright statement
%\IEEEpubidadjcol

\begin{abstract}
In recent years, the evolution of machine learning techniques has significantly impacted the field of intrusion detection, particularly within the context of the Internet of Things (IoT). As IoT networks expand, the need for robust security measures to counteract potential threats has become increasingly critical. This paper introduces a hybrid Intrusion Detection System (IDS) that synergistically combines Kolmogorov-Arnold Networks (KANs) with the XGBoost algorithm. Our proposed IDS leverages the unique capabilities of KANs, which utilize learnable activation functions to model complex relationships within data, alongside the powerful ensemble learning techniques of XGBoost, known for its high performance in classification tasks. This hybrid approach not only enhances the detection accuracy but also improves the interpretability of the model, making it suitable for dynamic and intricate IoT environments. Experimental evaluations demonstrate that our hybrid IDS achieves an impressive detection accuracy exceeding 99\% in distinguishing between benign and malicious activities. 
Additionally, we were able to achieve F1-scores, precision, and recall that are exceeding~98\%.
Furthermore, we conduct a comparative analysis against traditional Multi-Layer Perceptron (MLP) networks, assessing performance metrics such as Precision, Recall, and F1-score. The results underscore the efficacy of integrating KANs with XGBoost, highlighting the potential of this innovative approach to significantly strengthen the security framework of IoT networks. 
\end{abstract}


% \begin{figure}[h]
% \centering
% \includegraphics[width=80mm]{Figure1.png}
% \caption{Example of a figure caption.}
% \label{figure1}
% \end{figure}

\section{Introduction}\label{section1}

The Internet of Things (IoT) has come a long way since Kevin Ashton first introduced the term in 1999 \cite{nauman2020multimedia}. Initially, the idea was to connect RFID devices to the internet, but today, IoT has expanded to include a wide variety of electronic devices—ranging from sensors and actuators to smartphones and smart home appliances. These devices communicate with one another using unique identifiers, working together to accomplish various tasks. However, this integration of diverse devices also brings significant security challenges \cite{kocakulak2017overview}.

Although encryption is a fundamental aspect of IoT security and serves as the first line of defense against potential threats, relying solely on it is insufficient \cite{iacovazzi2023towards, butun2024expandable}. To effectively detect and respond to attacks, additional security measures, such as Location Obfuscation as shown in \cite{butun2019preserving}, and/or Intrusion Detection Systems (IDS) \cite{amouri2020machine, 8363921}, are essential. One significant threat to IoT networks is botnets—malicious software that can exploit poorly configured devices \cite{butun2019security}. To develop effective IDS solutions and address the unique security challenges faced by IoT systems, realistic and specifically designed datasets are crucial. Despite the availability of various cybersecurity datasets, there is a critical need for those tailored to the IoT environment.

To address these security issues, researchers have developed a range of Intrusion Detection Systems (IDS) that leverage machine learning methods. Combining Multilayer Perceptron (MLP) and chaotic neural networks, Shettar et al.~\cite{shettar2021intrusion} developed a hybrid model to enhance attack detection accuracy and precision while significantly decreasing false alarms. Their experiments using the KDD Cup '99 dataset showed that the hybrid model outperformed the MLP alone in reducing false alarms without sacrificing accuracy, precision, or recall. 

Mushtaq et. al.~\cite{mushtaq2022two} proposed an intrusion detection system, SE-IDS, which employs a stacked ensemble architecture. The base learners comprising the system include decision trees, XGBoost, bagging, extra trees, and random forest. The predictions generated by these base learners are subsequently fed into a multilayer perceptron (MLP) as a meta-learner to produce the final classification. The performance of the proposed framework is evaluated using the NSL-KDD dataset. 

A Multilayer Perceptron (MLP) network that utilizes a hybrid feature selection scheme namely, Information Gain Random Forest (IGRF) and Recursive Feature Elimination (RFE), is proposed by Yin et al.~\cite{yin2023igrf}. According to the authors, the scheme combines the efficiency of filter methods with the effectiveness of wrapper approaches for feature selection. The performance of the proposed scheme is evaluated using the UNSW-NB15 dataset. 

% Khater et al.~\cite{khater2019lightweight} introduced a lightweight IDS utilizing a vector space representation through a Multilayer Perceptron (MLP) model. The performance of this IDS was assessed using the Australian Defense Force Academy Linux Dataset (ADFA-LD) and the Australian Defense Force Academy Windows Dataset (ADFA-WD), which are contemporary datasets that incorporate various exploits and attacks on applications. The system's performance was tested on a Raspberry Pi, highlighting its suitability for resource-constrained environments.

Khater et al.~\cite{khater2019lightweight} presented a lightweight IDS based on a vector space representation using a Multilayer Perceptron model. This light IDS was evaluated against two relatively recent datasets that include different exploits and attacks against applications, like the Australian Defense Force Academy Linux Dataset and the Australian Defense Force Academy Windows Dataset. Performance evaluation regarding this system was done on a Raspberry Pi, thus showing its applicability to resource-constrained devices.

% HADMLP-IDS model for intrusion detection was introduced by Ghanem et al.~\cite{ghanem2020efficient}, which combines the Artificial Bee Colony and Dragonfly algorithms to train a Multilayer Perceptron. The model's effectiveness was evaluated using a confusion matrix and tested on several well-known datasets for intrusion detection, including KDD Cup 99, NSL-KDD, UNSW-NB15, and ISCX2012.

Ghanem et al.~\cite{ghanem2020efficient} proposed an intrusion detection model called HADMLP-IDS that combines the Artificial Bee Colony with Dragonfly algorithms for the training of a Multilayer Perceptron. The confusion matrix was used to evaluate the efficiency of this model and test its application on various well-known datasets for intrusion detection, including KDD Cup 99, NSL-KDD, UNSW-NB15, and ISCX2012.

In the fast-changing area of Intrusion Detection Systems (IDS) for IoT environments, it is essential to localize new and effective approaches that could enhance security. This paper proposes a state-of-the-art framework for an IDS that incorporates Kolmogorov-Arnold Networks (KANs)~\cite{liu2024kan} with XGBoost in a hybrid approach. To the best of the available literature, this would be the first application of KANs in intrusion detection tasks. The KAN component skillfully discovers hidden patterns in the data, and the XGBoost, underpinning it with fast training and high performance, makes it quite effective in classifying and detecting anomalies in network traffic. This kind of interaction brings out a very strong IDS that can rightly interpret network behavior to detect even small deviations. 

In the experiments, we used the N-BaIoT dataset \cite{meidan2018nbaio}, which contains two well-known Distributed Denial of Service (DDoS) botnet attacks: Mirai and Bashlite. By making KAN an integral part of our ensemble model, we solve some of the most complex challenges in the detection of sophisticated attacks within IoT environments, hence setting a new benchmark in the area with a comprehensive at the same time efficient solution for intrusion detection.

The remainder of this paper is structured as follows: Section~\ref{section2} provides a brief discussion on the KANs used to develop the IDS. Section~\ref{section3} describes the experimental setup and system architecture.Section~\ref{section4} presents the results and discussion. Finally, Section~\ref{section5} concludes the paper.

\section{Proposed Method }\label{section2}
%Our proposed Intrusion Detection System (IDS) leverages KANs as its core component. Unlike traditional MLPs, KANs employ a novel architecture inspired by the Kolmogorov-Arnold representation theorem. This section provides a foundational overview of KANs and their underlying principles.
At the heart of our proposed Intrusion Detection System (IDS) lie KANs. Unlike traditional MLPs, KANs are based on a radically new architecture inspired by the Kolmogorov-Arnold representation theorem. This section provides basic information about KANs and the basic approach behind them.

Our work is different in its new application to the Kolmogorov-Arnold Network within intrusion detection systems. This work presents, as a first study on KAN for intrusion detection systems, a well-explained and detailed analysis of how this improved network architecture can further enhance detection capabilities in IoT environments. This pioneering application not only manifests the versatility of KAN for a totally different domain but also charts a new way for further research into methodologies of intrusion detection systems.

% Introduced by [8], KANs distinguish themselves from MLPs using learnable activation functions on network edges, enabling more flexible and potentially powerful representations. Essentially, KANs process input values through a series of transformations involving univariate functions applied to network edges. This process is grounded in the Kolmogorov-Arnold representation theorem, which states that any multivariate function can be expressed as follows (refer to Equation~\ref{equ1}):

Introduced by [8], KANs differ from MPLs, in that they employ learnable activation functions at the network edges to allow a more flexible, and hence potentially powerful representation. More specifically, KANs work as a series of transformations of the input values described by univariate functions applied to the network edges. That is implemented following the Kolmogorov-Arnold representation theorem, which defines multivariate functions as the following formulation (see Eq. \ref{equ1}):

\begin{equation}\label{equ1}
f(x_{1}, \ldots, x_{l}) = \sum_{q=1}^{2l+1} \Phi_q \sum_{p=1}^l \phi_{q,p}(x_p)
\end{equation}

Where $\phi_{q,p}: [0,1] \rightarrow \mathbb{R}$ and $\Phi_q: \mathbb{R} \rightarrow \mathbb{R}$.

The structure of KANs organizes each layer as a matrix of these learnable 1D functions as follows (refer to Equation~\ref{equ2}):
% \[
\begin{equation}\label{equ2}
\Phi = \{\phi_{q,p}\}, \quad p = 1, \ldots, n_{in}, \quad q = 1, 2, \ldots, n_{out}
% \]
\end{equation}

% Each function $\phi_{q,p}$ can be defined as a B-spline, a type of spline function created by combining basis splines. This enhances the network's ability to learn intricate data patterns effectively. The number of input features to a particular layer is represented by $n_{in}$, while $n_{out}$ represents the number of output features produced by that layer, reflecting the dimensionality transformations across the network layers.

% Structurally, KANs resemble MLPs in their layered architecture, but they diverge by employing sophisticated functional mappings instead of simple linear transformations and fixed activations. A typical KAN comprises $M$ layers as follows (refer to Equation~\ref{equ3}):

Each activation function $\phi_{q,p}$ itself is an instance of B-splines, a special form of spline function acquired through the superposition of basis splines. This makes the network much more capable of learning complex patterns in the data. The number of features in input to a layer is $n_{in}$, and the number of features in the output generated by the same layer is $n_{out}$, indicating dimensionality transformations over the network layers.

Structurally, KANs are similar to MLPs in that they are layer-oriented. Still, in their functional mappings, the KANs utilize more complex mechanisms than just linear transformations or fixed activations. A typical KAN consists of $M$ layers, as in Eq. \ref{equ3}:
\begin{equation}\label{equ3}
% \[
\text{KAN}(x) = (\Phi_{M-1} \circ \Phi_{M-2} \circ \ldots \circ \Phi_1 \circ \Phi_0)x
% \]
\end{equation}

% The basic architecture of a KAN is illustrated in Figure~\ref{figure1}. Unlike traditional neural networks, KANs excel at capturing intricate data relationships due to their unique structure. By harnessing this capability, our proposed IDS aims to significantly enhance intrusion detection accuracy and efficiency.

\begin{figure}[!h]
\centering
\includegraphics[width=0.99\columnwidth]{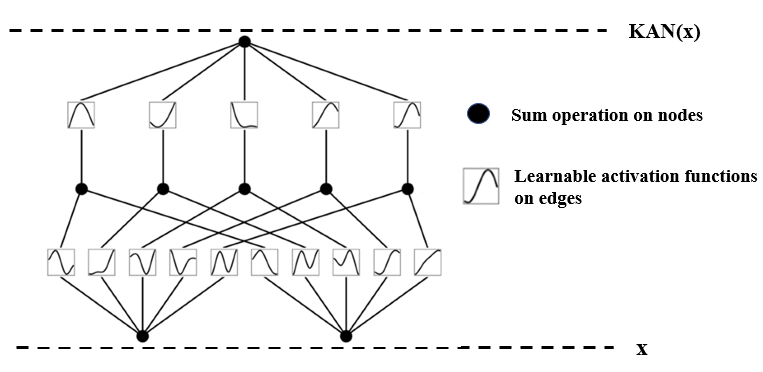}
\caption{A basic structure for KANS \cite{liu2024kan}}
\label{figure1}
\end{figure}

The overall architecture of KAN is shown in Figure~\ref{figure1}. Considering the unique structure, KANs are capable of capturing complicated relationships among data that traditional neural networks cannot. Our proposed IDS harnesses this capability to improve the accuracy and efficiency of intrusion detection significantly.

\begin{figure*}[!h]
\centering
\includegraphics[width=0.99\textwidth]{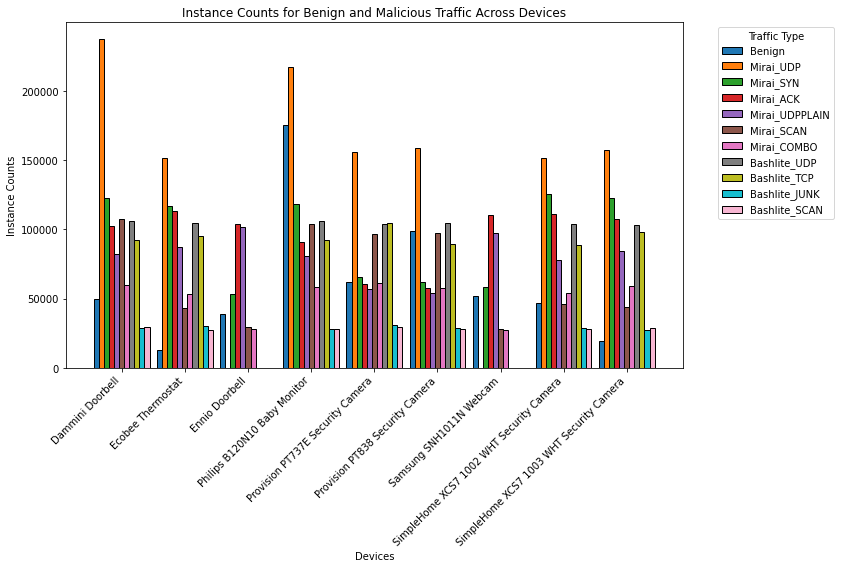}
\caption{A summary of the original N-BaIoT dataset instances per device(dataset) and per attack (class).}
\label{figure2}
\end{figure*}

\section{Experimental Results}\label{section3}
In this section, we describe the dataset used in our proposed model. Afterward, the experimental setting, evaluation metrics, and performance results have been discussed in detail.

\subsection{Dataset Description}

% Our study employs the N-BaIoT dataset, as introduced by Meidan \textit{et al.}~\cite{meidan2018nbaio}, which includes both benign and malicious network traffic. Malicious traffic was generated by injecting Bashlite and/or Mirai Botnet attacks into the network. The data was collected from nine common household devices, categorized into five types: thermostat, doorbell, webcam, security camera, and baby monitor. In the original dataset, the proportion of benign instances to total malicious instances varies from 1.59\% to 19\% for each device.

Our study is based on the N-BaIoT dataset, introduced by Meidan \textit{et al.}~\cite{meidan2018nbaio}, which includes both benign and malicious network traffic. Malicious traffic was generated by infecting the network with instances of the Bashlite and/or Mirai Botnet attacks. The data collection took place from nine different but commonplace devices representing five different device types: thermostat, doorbell, webcam, security camera, and baby monitor. In the original dataset, the ratio of benign cases against total malicious cases goes as low as 1.59\% to 19\% per device.

The dataset comprises 23 primary attributes, each measured across five different time frames with decay factors (lambda) of L5, L3, L1, L0.1, and L0.01. This results in a total of 115 features (23 x 5) for each instance. This dataset displays a diverse distribution of traffic types, including multiple Mirai and Bashlite attack variants. Each device’s dataset contains a substantial number of both benign and malicious instances, providing a comprehensive basis for evaluating the performance of the proposed IDS. This distribution is illustrated in Figure~\ref{figure2}.

% \begin{figure*}[!h]
% \centering
% \includegraphics[width=0.99\textwidth]{FIG2.png}
% \caption{A summary of the original N-BaIoT dataset instances per device(dataset) and per attack (class).}
% \label{figure2}
% \end{figure*}

For our experiments, we employ a reduced version of the original dataset, focusing on seven of the nine devices that were subjected to both Mirai and Bashlite attacks. The specific devices included in our study are as follows: Danmini doorbell, Ecobee thermostat, Phillips B120N10 Baby Monitor, Provision PT737E security camera, Provision PT838 security camera, Simple Home XCS7 1002 WHT security camera, Simple Home XCS7 1003 WHT security camera.

% Our experimental dataset consists of 500,000 instances, as shown in Figure~\ref{figure3}, and is organized as follows: 430,000 benign instances (Class 0) were randomly selected from seven device datasets, while 70,000 malicious instances (Classes 1-10) were generated by randomly sampling 1,000 samples from each type of attack within each device dataset.
% This balanced construction ensures a thorough representation of both normal and malicious network activities, offering a reliable foundation for training and evaluating our proposed Intrusion Detection System.

\begin{figure*}[h]
\centering
\includegraphics[width=0.99\textwidth]{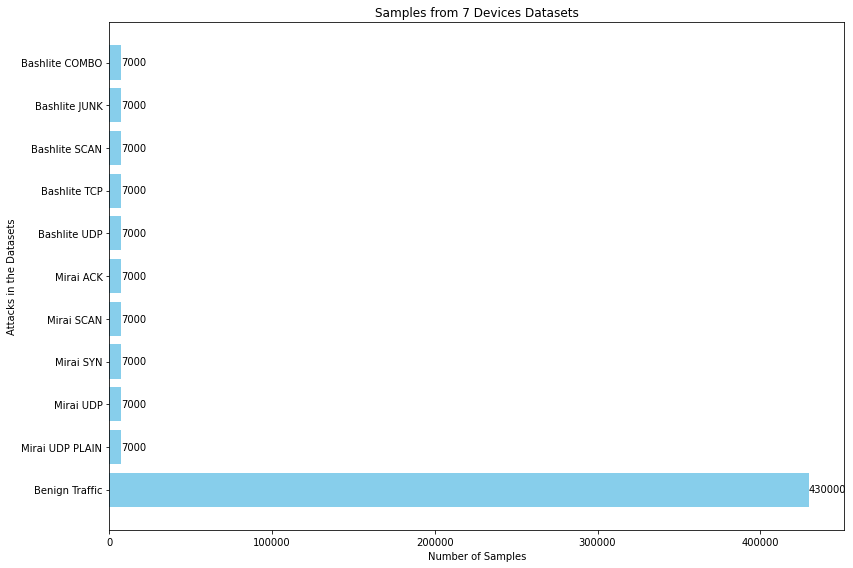}
\caption{Number of samples per class (Benign or Malicious) used in the experiments' dataset}
\label{figure3}
\end{figure*}

Our experimental dataset has 500,000 instances, shown in Figure~\ref{figure3}, structured as follows: 430,000 normal instances (Class 0) were randomly drawn from seven device datasets, and another 70,000 malicious instances (Classes 1-10) were prepared by randomly sampling 1,000 samples from every kind of attack for each device dataset.
This balanced construction will ensure that normal and malicious network activities are appropriately represented, which will establish a reliable base for training and testing our proposed Intrusion Detection System.

% \begin{figure}[!h]
% \centering
% \includegraphics[width=0.99\columnwidth]{FIG2.png}
% \caption{...}
% \label{figure2}
% \end{figure}

% \begin{figure}[!h]
% \centering
% \includegraphics[width=0.99\columnwidth]{FIG3.png}
% \caption{Number of samples per class (Benign or Malicious) used in the experiments' dataset}
% \label{figure3}
% \end{figure}

% \begin{figure*}[!h]
% \centering
% \includegraphics[width=0.59\textwidth]{FIG3.png}
% \caption{Number of samples per class (Benign or Malicious) used in the experiments' dataset}
% \label{figure3}
% \end{figure*}

\subsection{Experimental Setup}
% The proposed IDS employs a hybrid scheme that leverages both XGBoost and KAN networks. KANs can achieve high accuracy in modeling complex, non-linear relationships between inputs and outputs, while XGBoost is efficient and effective in handling large datasets with high performance. The feature space consists of 115 features, which serve as input to the KAN section. The output from the KAN is then fed into the XGBoost in a sequential manner. In this experiment, several hyperparameters were tuned to optimize the model's performance. The training process was conducted over 50 epochs with a batch size of 512. The learning rate was set to 0.001. The KAN used in the model comprised a single layer with 10 output units. The input features numbered 115, and the model was designed to classify 11 output classes. The KAN interval and order were set to 7 and 5, respectively. The cross-entropy loss function was employed, and the ADAM optimizer was utilized for optimization. A StepLR scheduler was applied with a step size of 10 and a gamma of 0.5 to adjust the learning rate. For the XGBoost component of the model, the number of estimators was set to 100, the learning rate was 0.1, and the maximum depth was configured to 6.

The proposed IDS uses a hybrid scheme that includes both XGBoost and KANs. On the other hand, in the case of complex, nonlinear relationships between the inputs and outputs, KANs will be very accurate. XGBoost will deal with large datasets efficiently at high performance. The feature space is composed of 115 features that will feed the KAN section. The output from KAN is fed into the XGBoost sequentially. In this experiment, quite a good number of hyper-parameters were tuned to best suit the model. In this context, training was made for 50 epochs with a batch size of 512. The learning rate used in this study is 0.001. The KAN applied in the model consisted of a single layer with 10 output units. The input features were 115, while the model was designed for classification into 11 output classes. The KAN interval and order were set to 7 and 5, respectively. In this regard, the cross-entropy loss function was used, while optimization utilized the ADAM optimizer. A StepLR scheduler was applied with a step size of 10 and gamma of 0.5 for adjusting the learning rate. For the component XGBoost, a number of estimators was set to 100, with the learning rate set to 0.1 and the maximum depth configured to 6.

\section{Results and Discussions}\label{section4}
We analyze the performance of our proposed IDS scheme using key metrics: Precision, Recall, F1 score, and Accuracy. The results presented in Table~\ref{table1} demonstrate that our hybrid KAN-XGBoost algorithm surpasses both MLP and KAN across all evaluated metrics. Notably, KAN-XGBoost achieves the highest scores with an accuracy of 99.69\%, precision of 98.1\%, recall of 98.01\%, and F1-score of 98.04\%. While KAN alone outperforms MLP, the integration of XGBoost with KAN significantly boosts overall performance, highlighting the hybrid model's capacity to enhance function approximation and effectively handle large datasets.

Here, performance metrics are computed to evaluate the proposed IDS scheme. Table~\ref{table1} illustrates that our proposed hybrid KAN-XGBoost algorithm outperforms both MLP and KAN algorithms across all the evaluated metrics. Notably, KAN-XGBoost produces an accuracy of 99.69\%, a precision of 98.1\%, a recall of 98.01\%, and an F1-score of 98.04\%. While KAN alone outperforms MLP, it is when combining XGBoost with KAN that substantial performance improvements are realized, thus underpinning the capability of a hybrid model toward improved function approximation and dealing with large datasets.

\begin{table}[!h]
\centering
\caption{Results comparison between MLP, KAN, and the proposed scheme}
\begin{tabular}{lcccc}
\toprule
Algorithm & Accuracy & Precision & Recall & F1-score \\
name & (\%) & (\%) & (\%) & (\%) \\
\midrule
MLP & 96.46 & 95.75 & 96.46 & 95.97 \\
KAN & 97.37 & 96.77 & 97.37 & 96.87 \\
Proposed &  &  &  &  \\
KAN-XGBoost & 99.69 & 98.10 & 98.01 & 98.04 \\
\bottomrule
\end{tabular} \label{table1}
\end{table}

\begin{figure*}[!h]
\centering
\includegraphics[width=0.85\textwidth]{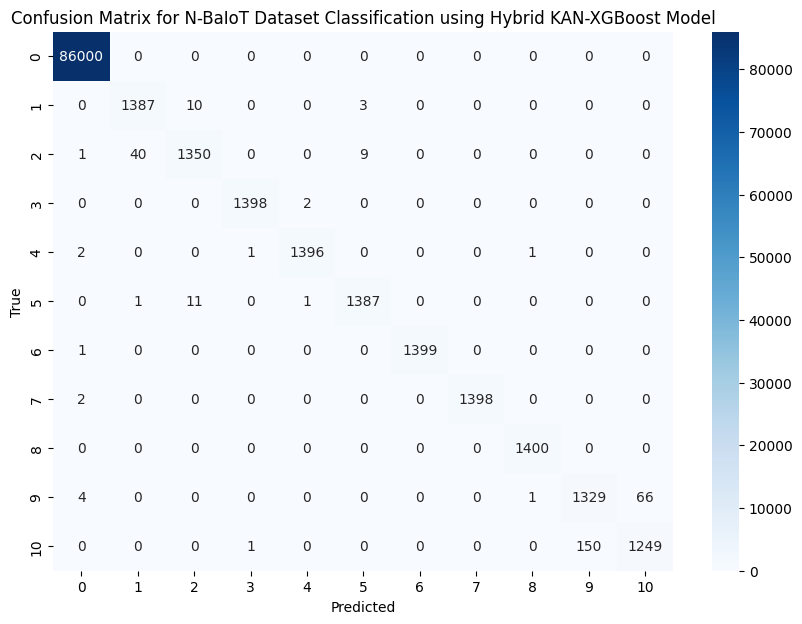}
\caption{Confusion matrix for the proposed KAN-XGBoost scheme}
\label{figure5}
\end{figure*}

The confusion matrix for the N-BaIoT dataset classification using the hybrid KAN-XGBoost model is shown in Figure~\ref{figure5}, highlighting the model's high accuracy and precision. Most classes are classified with near-perfect precision, demonstrating the model's effectiveness in distinguishing various types of network traffic. 

% The inspection of epoch versus loss for KAN, MLP, and KAN-XGBoost models reveals that all three models exhibit significant loss reduction in the initial epochs, with KAN and MLP showing a more pronounced plateau after about 30 epochs, indicating early convergence as shown in Figure~\ref{figure4}. KAN-XGBoost, however, demonstrates a steady and continuous decrease in loss throughout the 50 epochs, achieving the lowest final loss among the models. This suggests that the hybrid KAN-XGBoost model benefits from the combined strengths of both techniques, resulting in a more persistent and effective learning curve, thereby potentially offering better performance and generalization compared to the individual models.

\begin{figure*}[h]
\centering
\includegraphics[width=0.99\textwidth]{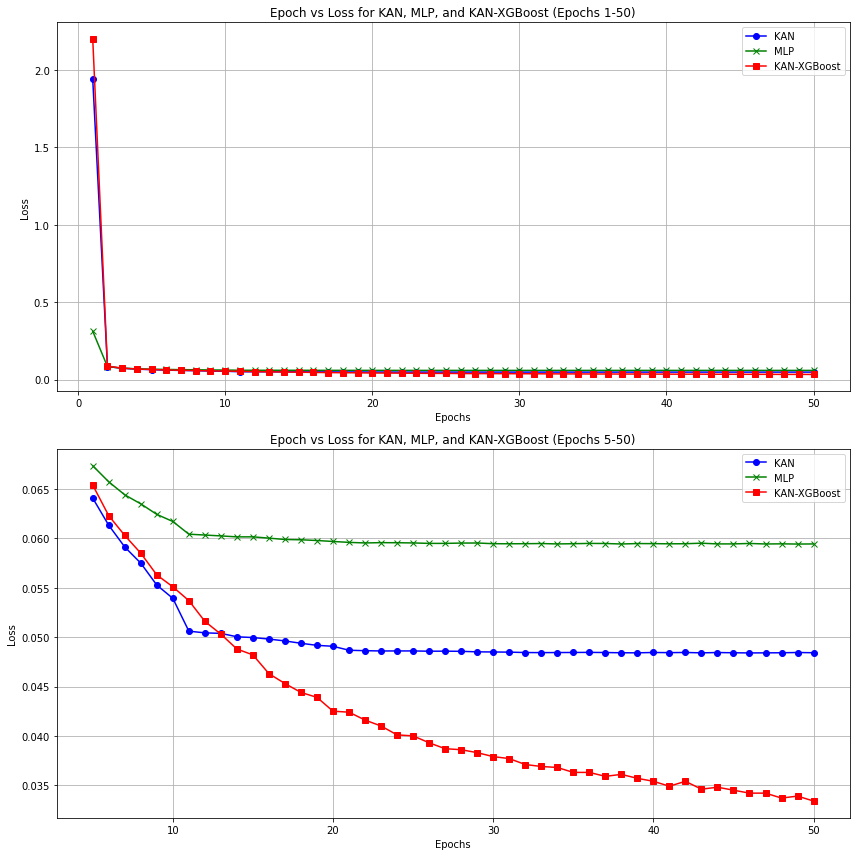}
\caption{Epoch count vs loss for KAN, MLP, and KAN-XGBoost, the lower part is from epoch 5 to epoch 50.}
\label{figure4}
\end{figure*}

The plot of the epoch vs. loss for the KAN, MLP, and KAN-XGBoost models shows all models exhibit dramatic losses in the early epochs; however, KAN and MLP plateau more extremely after around 30 epochs, thus showing early convergence, as depicted in Figure~\ref{figure4}. KAN-XGBoost showed a much more steady and continuous decrease in loss over the 50 epochs, with the lowest final loss of the three models. This means that the hybrid KAN-XGBoost model reaps the benefits of both techniques in such a way that it gives back a more persistent and efficient learning curve for better performance and generalization compared to individual models.

% \begin{figure*}[h]
% \centering
% \includegraphics[width=0.49\textwidth]{FIG5.png}
% \caption{Confusion matrix for the proposed KAN-XGBoost scheme}
% \label{figure5}
% \end{figure*}

%Importantly, both benign traffic and several attack classes are detected with minimal misclassifications, showcasing the robustness of the hybrid approach in handling diverse and complex intrusion scenarios. In conclusion, our analysis establishes a clear performance ranking: KAN-XGBoost emerges as the top performer, followed by KAN, with MLP demonstrating the least favourable performance. These findings underscore the effectiveness of our proposed hybrid approach in intrusion detection tasks.
It is concluded that benign traffic and many classes of attacks are detected with the least number of misclassifications, thus establishing the robustness of the hybrid approach at hand for different and complex intrusion scenarios. It is vital to note that our analysis very clearly ranks performance as follows: KAN-XGBoost is the top performer; next is KAN, and the least favorable is MLP. These results underscore the efficacy of our proposed hybrid approach to the tasks of intrusion detection.

% \begin{figure}[!h]
% \centering
% \includegraphics[width=0.99\columnwidth]{FIG4.png}
% \caption{Epoch count vs loss for KAN, MLP, and KAN-XGBoost, the lower part is from epoch 5 to epoch 50.}
% \label{figure4}
% \end{figure}

% \balance

\section{Conclusions}\label{section5}
% This paper introduces an innovative Intrusion Detection System (IDS) that leverages KANs in a hybrid KAN-XGBoost model. By combining KANs' high learnability through adaptive activation functions with XGBoost's efficiency in handling large datasets and capturing intricate patterns, the proposed system achieves excellent performance. Evaluated using the N-BaIoT dataset, the model outperforms standalone MLP and KAN networks, demonstrating strong metrics with an Accuracy of 99.69\%, Precision of 98.1\%, Recall of 98.01\%, and an F1-score of 98.04\%. 

% These results underscore the effectiveness of the hybrid approach in detecting network intrusions with high accuracy and reliability. Future research will focus on optimizing hyperparameters and exploring alternative KAN-based structures. This work represents a significant advancement in network security, providing a solid foundation for developing more effective defence mechanisms against evolving cyber threats.

The paper presents a new Intrusion Detection System based on KANs in a hybrid model, called KAN-XGBoost. In this model, the high learnability of KANs by means of adaptive activation functions is combined with the efficiency of XGBoost in dealing with large datasets and capturing complex patterns, making it an excellent performing model. The model evaluated on the N-BaIoT dataset outperforms standalone MLP and KAN networks with very strong metrics in accuracy, precision, recall, and F1-score of 99.69\%, 98.1\%, 98.01\%, and 98.04\%, respectively. 

These results underpin the effectiveness of the hybrid approach in detecting network intrusions with a high degree of accuracy and reliability. Future research will then be focused on the optimization of hyperparameters and looking at other structures that are KAN-based. This contribution greatly enriches the network security field, hence offering a rather solid background for the development of more effective defense mechanisms against ever-evolving cyber threats.

% \newpage
%\vspace{0.65cm}
% \section*{Acknowledgement}
% This work is done in the Tecore Networks Laboratory at Florida Atlantic University (FAU) under the department of EECS and is funded by the Office of the Secretary of Defense (OSD), Grant Number W911NF2010300.

% \vspace{2.65cm}

%\newpage
\section*{Appendix}
The abbreviations used within this text are summarized in Table~\ref{table2}:

\begin{table} [!h]
\caption{Abbreviation table} \label{table2}
\centering
%% \tablesize{} %% You can specify the fontsize here, e.g.  \tablesize{\footnotesize}. If commented out \small will be used.
\begin{tabular}{|>{\raggedright}m{2cm}|!{\raggedright}m{5cm}|}
% \hline
\hline
\rowcolor{lightgray}
\hline
\textbf{Abbreviation}&\textbf{Description }\\
\hline
\hline
\rowcolor{white}
\hline
DDoS &  Distributed Denial of Service \\
\hline
\hline
\rowcolor{lightgray}
\hline
IDS &  Intrusion Detection Systems\\
\hline
\hline
\rowcolor{white}
\hline
IGRF & Information Gain Random Forest\\ 

\hline
\hline
\rowcolor{lightgray}
\hline
IoT & Internet of Things\\

 \hline
\hline
\rowcolor{white}
\hline
KANs & Kolmogorov-Arnold Networks \\

\hline
\hline
\rowcolor{lightgray}
\hline
MLP & Multilayer Perceptron \\

 \hline
\hline
\rowcolor{white}
\hline
RFE & Recursive Feature Elimination\\

\hline
\hline
\rowcolor{lightgray}
\hline
XGBoost & Extreme Gradient Boosting \\

\hline
\end{tabular}
\end{table}

%\vspace{0.65cm}
 \newpage
 .
 \newpage
 .
\newpage

%\balance

\bibliographystyle{IEEEtran}
\bibliography{conference_101719.bib}

\end{document}